\documentclass[preprint,amsmath,amssymb,aps,pre,longbibliography]{revtex4-1}

\usepackage{amsmath}
\usepackage{amssymb}

\usepackage{graphicx}
\usepackage{dcolumn}
\usepackage{bm}
\usepackage{cancel}
\usepackage{color}
\usepackage[utf8]{inputenc}

\begin{document}

\title{Transition rates of non-interacting quantum particles from the Widom insertion formula}

\author{M. Hoyuelos}
\email{hoyuelos@mdp.edu.ar}
\affiliation{Instituto de Investigaciones F\'isicas de Mar del Plata (IFIMAR-CONICET), Departamento de F\'isica, Facultad de Ciencias Exactas y Naturales, Universidad Nacional de Mar del Plata, Funes 3350, 7600 Mar del Plata, Argentina}

\begin{abstract}
Transition rates among different states in a system of non-interacting quantum particles in contact with a heat reservoir include the factor $1\mp \bar{n}_i$, with a minus sign for fermions and a plus sign for bosons, where $\bar{n}_i$ is the average occupation number of the final state. It is shown that this factor can be related to the difference of the chemical potential from that of an ideal classical mixture; this difference is formally equivalent to the excess chemical potential in a classical system of interacting particles. Using this analogy, Widom's insertion formula is used in the calculation of transition rates. The result allows an alternative derivation of quantum statistics from the condition that transition rates depend only on the number of particles in the target energy level. Instead, if transition rates depend on the particle number only in the origin level, the statistics of ewkons is obtained; this is an exotic statistics that can be applied to the description of dark energy.
\end{abstract}


\maketitle

\section{Introduction}

Rate equations, that govern the time evolution of average occupation numbers $\bar{n}_i$ of quantum states with energy $\epsilon_i$ of a system of particles in contact with a reservoir at temperature $T$, have important applications in the description of, for example, Fermi gas, phonons or Bose-Einstein condensation \cite{banyai2}, see also \cite{brenig,banyai}. They have the form of the master equation,
\begin{equation}\label{e.master}
\frac{\partial \bar{n}_i}{\partial t} = \sum_j (\bar{n}_j W_{\bar{n}_j,\bar{n}_i}  - \bar{n}_i W_{\bar{n}_i,\bar{n}_j})
\end{equation}
with transition rates given by
\begin{equation}\label{e.transrates}
W_{\bar{n}_j,\bar{n}_i} = \nu e^{-\beta (\epsilon_i-\epsilon_j)/2} (1\mp \bar{n}_i)
\end{equation}
where the minus sign is for fermions and the plus sign is for bosons, and $\nu$ is a jump frequency; $W_{\bar{n}_j,\bar{n}_i}$ is the transition rate per particle in state $j$ from state $j$ to state $i$; in both cases $W_{\bar{n}_j,\bar{n}_i}$ depends on the occupation number in the destination state, $\bar{n}_i$, but does not depend on $\bar{n}_j$. The factor $1\mp \bar{n}_i$ is generally justified using heuristic arguments, see \cite[Ch.\ 32]{brenig} or \cite[Sec.\ 8.2 and 21.2]{banyai}. For example, citing \cite{brenig}, ``in Fermi gases and liquids processes can occur only if the final state is unoccupied'', and factor $1-\bar{n}_i$ is included; there is not an equivalent argument to justify factor $1+\bar{n}_i$ for bosons. A similar procedure is followed in \cite{banyai}. These rates reproduce the correct average occupation numbers in equilibrium, that is, the Fermi-Dirac and Bose-Einstein distributions (see the Appendix), and this result can be considered as a sufficient (heuristic) argument for factors $1\mp \bar{n}_i$.

Transition rates proportional to $1\mp \bar{n}_i$ are used in \cite{kania} to obtain a Fokker-Planck equation for the occupation number in a continuous energy space, see also \cite{kania0,kania2} and \cite[Sec.\ 6.8]{kadanoff}. Fokker-Planck equations for fermions and bosons can also be derived from the free energy functional, see \cite{frank0} and \cite[Sec.\ 6.5.4]{frank}; in particular, the so-called drift form of the Fokker-Planck equation involves a mobility that behaves as $1\mp \bar{n}_i$. A Langevin approach has been developed in \cite{frank1}. Fundamental applications are the metal electron and black body radiation models \cite[Sec.\ 6.5.6]{frank}. It can be considered that $\bar{n}_i$ represents the number of particles in a degenerate energy level $i$ that includes $g_i$ different states; in this case, the mentioned factors for fermions and bosons are $1\mp \bar{n}_i/g_i$.

In this paper, a method is proposed for the derivation of transition rates for quantum systems of non-interacting particles. It is based on a previous work for classical systems with interacting particles \cite{dimuro}. The main ingredients are the detailed balance relationship and the Widom insertion formula. From detailed balance, transition rates can be related with the insertion energy, that is, the energy needed to add one particle. And from the Widom insertion formula \cite{widom}, the insertion energy is related to the excess chemical potential. A separation of terms with different orders of an extensive quantity is used to finally obtain the transition rate in terms of the excess chemical potential, the energy difference and an undetermined jump frequency. In the case of non-interacting quantum particles that is analyzed here, the excess chemical potential, $\mu_{\text{q}}$, represents quantum effects instead of interactions. In both cases (classical with interactions or quantum without interactions), it gives the departure of the chemical potential from the classical ideal case. 

The main difference with previous approaches is that the method demonstrates a direct connection between transition rates and statistics. It provides a formula for transitions rates in terms of the excess chemical potential, that is directly obtained from the occupation number distributions for fermions or bosons.

The generalization to the quantum case of the derivation of transition rates $W_{\bar{n}_1,\bar{n}_2}$ between generic levels 1 and 2, developed in Ref.\ \cite{dimuro}, is presented in Sec.\ \ref{s.theory}. When the excess chemical potential for fermions or bosons is used in the result for the transition rate, Eq.\ \eqref{e.transrates} is obtained. In Sec.\ \ref{s.destlevel} it is shown that, before assuming a specific form for $\mu_{\text{q}}$, if the transition rate depends only on the particle number in the destination level, that is, if $W_{\bar{n}_1,\bar{n}_2}$ depends only on $\bar{n}_2$, then the known statistics of Bose-Einstein, Fermi-Dirac and Maxwell-Boltzmann are obtained. On the other hand, Sec.\ \ref{s.exotic} shows that if $W_{\bar{n}_1,\bar{n}_2}$ depends only on the occupation number in the origin level, $\bar{n}_1$, then the exotic statistics of ewkons is obtained; ewkons have been introduced in \cite{hoyuelos-sisterna} to describe dark energy. A summary and conclusions are presented in Sec.\ \ref{s.conclusions}

\section{Theory}
\label{s.theory}

Let us consider the quantum formulation of non-interacting particles in a volume $V$, in equilibrium with a reservoir at temperature $T$ and chemical potential $\mu$. There is an undetermined number of energy levels identified with index $i$. Each level, with energy $\epsilon_i$, has $n_i$ particles. Some basic concepts are reviewed in the next paragraphs to make the notation clear. 

\subsection{Basic formulae}
\label{s.basic}

The canonical partition function for $N$ particles is
\begin{equation}\label{eq:Z}
\mathcal{Z}_N = \text{tr } e^{-\beta \hat{H}} = {\sum_{\{n_i\}}}' \prod_i \mathcal{Z}_{n_i}
\end{equation}
where $\beta=(k_B T)^{-1}$ and $\hat{H}$ is the free Hamiltonian; see, for example, \cite[Chap.\ 12]{greiner} or \cite[Chap.\ 6]{pathria}. The sum $\sum_{\{n_i\}}'$ is over all $n_i$ that satisfy $\sum_i n_i = N$, and $\mathcal{Z}_{n_i}$ is the canonical partition function for level $i$. For fermions or bosons, we know that $\mathcal{Z}_{n_i}$ is equal to $e^{-\beta \epsilon_i n_i}$ and the set of possible values of $\{n_i\}$ has to be determined for each case. For classical particles,
\begin{equation}\label{eq:Zcl}
\mathcal{Z}_{n_i}^\text{cl} = e^{-\beta \epsilon_i n_i}/n_i!,
\end{equation}
and the canonical partition function of the whole system is $\mathcal{Z}_N^\text{cl} = (V/\lambda^3)^N/N!$,  where $\lambda$ is the thermal de Broglie wavelength \cite[p.\ 147]{pathria}.

The grand partition function is
\begin{equation}\label{eq:Q}
\mathcal{Q}(\mu,T,V) = \text{tr }e^{\beta(\mu \hat{N} - \hat{H})} = \sum_{N=0}^{\infty} \mathcal{Z}_N e^{\beta \mu N} = \prod_i \mathcal{Q}_i
\end{equation}
where $\hat{N}$ is the total number of particles operator and
\begin{equation}\label{eq:Qi}
\mathcal{Q}_i = \sum_{n_i} e^{\beta \mu n_i} \mathcal{Z}_{n_i}
\end{equation}
is the grand partition function for level $i$, where the sum is on all allowed values of $n_i$ (0 to $\infty$ for bosons or classical particles and 0 or 1 for fermions). Each element of the sum in Eq.\ \eqref{eq:Qi} is proportional to the probability $P_{n_i}$ of having $n_i$ particles:
\begin{equation}\label{eq:Pni}
P_{n_i} = \frac{e^{\beta \mu n_i} \mathcal{Z}_{n_i}}{\mathcal{Q}_i}.
\end{equation}
This probability is used below when introducing detailed balance and transition rates.

\subsection{Average number of particles and the Widom insertion formula}

If volume $V$ is large, single particle energy levels are very close to each other. It is considered that one level with energy $\epsilon_i$ actually encompasses a group of $g_i$ levels that are very close; number $g_i$ also includes a possible degeneracy. Then, the average number of particles with energy $\epsilon_i$ is given by
\begin{equation}\label{eq:nav}
\bar{n}_i = g_i \frac{1}{\beta}\frac{\partial \ln\mathcal{Q}_i}{\partial\mu}.
\end{equation}
Number $g_i$, and $\bar{n}_i$, are extensive quantities, proportional to $V$; $g_i$ is used later as an expansion parameter.

Different statistics are treated here in a unified way. In order to do that, the effective energy, $\phi_{n_i}$, defined as 
\begin{equation}\label{eq:effen}
e^{-\beta \phi_{n_i}} = \frac{\mathcal{Z}_{n_i}}{\mathcal{Z}_{n_i}^\text{cl}},
\end{equation}
is introduced as a parameter to measure non-classical effects: $\phi_{n_i}=0$ for classical particles, and it is different from zero if $\mathcal{Z}_{n_i}\ne \mathcal{Z}_{n_i}^\text{cl}$. For fermions, $\phi_{n_i}$ diverges if $n_i>1$.

Using Eqs.\ \eqref{eq:Qi}, \eqref{eq:effen} and \eqref{eq:Zcl} to calculate the average number of particles \eqref{eq:nav}, we get (subscript $i$ in $n_i$ can be removed from the sum index to simplify the notation at this stage)
\begin{align}
\frac{\bar{n}_i}{g_i} &= \frac{1}{\mathcal{Q}_i} \sum_{n=0}^\infty \frac{n}{n!} e^{-\beta(\epsilon_i-\mu)n} e^{-\beta\phi_{n}} \nonumber\\
&=\frac{e^{-\beta(\epsilon_i-\mu)}}{\mathcal{Q}_i} \sum_{n=1}^\infty \frac{1}{(n-1)!} e^{-\beta[\phi_n+(\epsilon_i-\mu)(n-1)]}  \nonumber\\
&=\frac{e^{-\beta(\epsilon_i-\mu)}}{\mathcal{Q}_i} \sum_{m=0}^\infty \frac{1}{m!} e^{-\beta(\phi_{m+1}-\phi_m)}  e^{-\beta[\phi_m+(\epsilon_i-\mu)m]} \nonumber\\
&= e^{-\beta(\epsilon_i-\mu)} \langle e^{-\beta \Delta\phi_n} \rangle,
\label{eq:barn}
\end{align}
with $\Delta\phi_n = \phi_{n+1}-\phi_n$. As expected, if $\phi_{n}=0$ then classical statistics is recovered. Now, we define $\mu_\text{q}$ as
\begin{equation}\label{eq:muq}
e^{-\beta \mu_\text{q}} = \langle e^{-\beta \Delta\phi_n} \rangle.
\end{equation}
It can be interpreted as a correction to the chemical potential representing non-classical effects; it is an intensive quantity of order $\mathcal{O}(g_i^0)$. We use the following notation: when $\mu_\text{q}$ is written without subscript $n$, it is evaluated at the average number of particles $\bar{n}_i$; instead, $\mu_{\text{q},n}$ is evaluated at a specific number of particles $n$. Its derivative respect to the number of particles, $\mu_{\text{q},n}'= \frac{\partial \mu_{\text{q},n}}{\partial n}$, is $\mathcal{O}(g_i^{-1})$.

From the average particle number in each case, it is known that
\begin{equation}\label{eq:stat}
e^{-\beta \mu_\text{q}}= \left\{
\begin{array}{ll}
1 & \text{for classical particles,} \\
\frac{1}{1+e^{-\beta(\epsilon_i-\mu)}} = 1-\bar{n}_i/g_i & \text{for fermions,} \\
\frac{1}{1-e^{-\beta(\epsilon_i-\mu)}} = 1+\bar{n}_i/g_i & \text{for bosons.}
\end{array}\right.
\end{equation}
We will use $\mu_\text{q}$ in order to treat all possible statistics in a unified way, as mentioned before. Eq.\ \eqref{eq:muq} is formally equivalent to the Widom insertion formula \cite{widom} (see also \cite[p.\ 30]{hansen2}) since $\mu_\text{q}$ is interpreted as the excess chemical potential and $\Delta\phi_{n}$ is interpreted as the insertion energy (the interaction energy needed to insert one particle) in a classical system of interacting particles. In the next subsection, the differences $\phi_{n_2+1}-\phi_{n_2}$ and $\phi_{n_1}-\phi_{n_1-1}$, for energy levels 1 and 2, are needed. They can be written in terms of $\mu_{\text{q},n_i}$ using \eqref{eq:muq} and following the same formal steps that are described in Ref.\ \cite{dimuro}:
\begin{align}
\phi_{n_2+1}-\phi_{n_2} &= \mu_{\text{q},n_2} - \frac{1}{2\beta} \frac{\Gamma'_{n_2}}{\Gamma_{n_2}} + \frac{\mu'_{\text{q},n_2}}{2}  + \text{h.t.} \label{eq:Dphi2} \\
\phi_{n_1}-\phi_{n_1-1} &= \underbrace{\vphantom{\frac{1}{\Gamma_{n_2}}}\mu_{\text{q},n_1}}_{\mathcal{O}(1)} - \underbrace{ \frac{1}{2\beta} \frac{\Gamma'_{n_1}}{\Gamma_{n_1}} - \frac{\mu'_{\text{q},n_1}}{2}}_{\mathcal{O}(g_i^{-1})} +\; \text{h.t.} \label{eq:Dphi1}
\end{align}
where $\Gamma_{n_i} = 1 + \beta n_i \mu_{\text{q},n_i}'$, and higher order terms of $1/g_i$ are represented by ``h.t.''.

\subsection{Transition rates}

Let us use labels 1 and 2 for two generic energy levels in an initial state with $n_1$ and $n_2$ particles. After a jump of one particle from level 1 to level 2, the final state has $n_1-1$ and $n_2+1$ particles. Let us call $W_{AB}$ the transition rate from state $A=\{n_1,n_2\}$ to state $B=\{n_1-1,n_2+1\}$, and $W_{BA}$ the rate of the inverse process. The system is in equilibrium and detailed balance is satisfied:
\begin{equation}\label{eq:detbalall}
P_A\, W_{A,B} = P_B\, W_{B,A},
\end{equation}
with $P_A$ and $P_B$ the probabilities of states $A$ and $B$. These probabilities are $P_A = P_{n_1} P_{n_2}$ and $P_B = P_{n_1-1} P_{n_2+1}$ with $P_{n_i}$ given by \eqref{eq:Pni}.
Then,
\begin{equation}\label{eq:detbala2}
\mathcal{Z}_{n_1}\mathcal{Z}_{n_2}\, W_{A,B} = \mathcal{Z}_{n_1-1}\mathcal{Z}_{n_2+1}\, W_{B,A},
\end{equation}

From Eqs.\ \eqref{eq:detbala2} and \eqref{eq:effen} we have
\begin{equation}\label{eq:ww}
\frac{W_{A,B}}{W_{B,A}} = \frac{e^{-\beta(\phi_{n_2+1}-\phi_{n_2})}}{e^{-\beta(\phi_{n_1}-\phi_{n_1-1})}} \frac{\mathcal{Z}_{n_1-1}^\text{cl} \mathcal{Z}_{n_2+1}^\text{cl}}{\mathcal{Z}_{n_1}^\text{cl} \mathcal{Z}_{n_2}^\text{cl}} = \frac{e^{-\beta(\phi_{n_2+1}-\phi_{n_2})}}{e^{-\beta(\phi_{n_1}-\phi_{n_1-1})}} \frac{n_1}{n_2+1} e^{-\beta(\epsilon_2-\epsilon_1)},
\end{equation}
where Eq.\ \eqref{eq:Zcl} for $\mathcal{Z}_{n_i}^\text{cl}$ was used.

The rate $W_{A,B}$ corresponds to the jump of any of the $n_1$ particles from level 1 to level 2. The transition rates per particle in the origin level are  $W_{n_1,n_2} = W_{A,B}/n_1$, with $n_1\ne 0$, and $W_{n_2+1,n_1-1} = W_{B,A}/(n_2+1)$, where the order of subindices indicates the jump direction. Then,
\begin{equation}\label{eq:wwt}
W_{n_1,n_2}\, e^{-\beta(\epsilon_1+\phi_{n_1}-\phi_{n_1-1})}
= W_{n_2+1,n_1-1}\, e^{-\beta(\epsilon_2+\phi_{n_2+1}-\phi_{n_2})}.
\end{equation}
The next step is to use Eqs.\ \eqref{eq:Dphi2} and \eqref{eq:Dphi1} for $\phi_{n_2+1}-\phi_{n_2}$ and $\phi_{n_1}-\phi_{n_1-1}$ and to separate different orders. Orders $g_1^{-1}$ and $g_2^{-1}$ are equivalent since $g_1$ and $g_2$ are proportional to $V$. Details of these calculations are described in Ref.\ \cite{dimuro} in the context of interacting classical particles, where the excess chemical potential, $\mu_{\text{ex},n_i}$, has to be replaced by $\mu_{\text{q},n_i}+\epsilon_i$. The result is
\begin{equation}\label{eq:w}
W_{n_1,n_2} = \nu e^{-\beta(\epsilon_2-\epsilon_1)/2} \frac{e^{\beta \mu_{\text{q},n_1}/2}}{(1 + \beta n_1 \mu_{\text{q},n_1}')^{1/2}} \frac{e^{-\beta \mu_{\text{q},n_2}/2}}{(1 + \beta n_2 \mu_{\text{q},n_2}')^{1/2}}.
\end{equation}
where $\nu$ is, in general, a function of the sum $n_1+n_2$; it represents the strength of the coupling with the heat reservoir and determines the time scale of the transition rate. This general approach is useful to determine the form of transition rates associated to different statistics, that are represented by $\mu_{\text{q},n_i}$, but leaves factor $\nu$ undetermined.

Using \eqref{eq:stat} for $e^{-\beta \mu_{\text{q},n_i}}$ in \eqref{eq:w}, we get the transition rates for the known statistics,
\begin{equation}\label{eq:ws}
W_{n_1,n_2} = \left\{ \begin{array}{ll}
\nu e^{-\beta(\epsilon_2-\epsilon_1)/2} & \text{classical particles} \\
\nu e^{-\beta(\epsilon_2-\epsilon_1)/2} (1-n_2/g_2) & \text{fermions} \\
\nu e^{-\beta(\epsilon_2-\epsilon_1)/2} (1+n_2/g_2) & \text{bosons.}
\end{array} \right.
\end{equation}
The result for fermions has an immediate interpretation: the transition rate is proportional to the number of states available in the destination energy level, $g_2-n_2$, due to the Pauli exclusion principle. As mentioned in the introduction, these transition rates for non-interacting quantum particles can be found in the literature, usually derived through heuristic arguments; see for example \cite[Ch.\ 32]{brenig} and \cite[Sec.\ 8.2 and 21.2]{banyai}. Reproduction of the known results confirms the validity of the procedure leading to Eq.\ \eqref{eq:w}. 


\section{Dependence on destination level}
\label{s.destlevel}

From \eqref{eq:ws}, it can be seen that transition rates depend only on the number of particles in the destination level, $n_2$. If the condition that $W_{\bar{n}_1,\bar{n}_2}$ depends only on $\bar{n}_2$ is assumed, then statistics of fermions, bosons, and classical particles should be obtained. This assertion is verified as follows. Considering that Eq.\ \eqref{eq:w} is evaluated at the average particle numbers, the condition implies that the factor that depends on $\bar{n}_1$ must be constant, that is
\begin{equation}\label{eq:const}
\frac{e^{\beta \mu_{\text{q}}/2}}{(1 + \beta \bar{n} \mu_{\text{q}}')^{1/2}} = 1,
\end{equation}
where subindex 1 is removed in $\bar{n}_1$ to lighten the notation. The constant on the right side is set equal to 1 taking into account the case of classical particles, for which $\mu_{\text{q}}=0$. Solutions of this equation are given by
\begin{equation}\label{eq:sols}
e^{-\beta \mu_{\text{q}}} = 1 + \kappa \bar{n},
\end{equation}
where $\kappa$ is a constant. Comparing with \eqref{eq:stat}, $\kappa=0$ corresponds to classical particles, $\kappa=-1/g$ corresponds to fermions and $\kappa=1/g$ corresponds to bosons (subindex 1 is removed in $g_1$ for simplicity).

\section{Exotic statistics}
\label{s.exotic}

The previous analysis, in which it has been shown that the assumption of transition rates depending only on $\bar{n}_2$ is enough to derive known statistics, suggests the possibility of exotic statistics that may be deduced from the theory when time is reversed. The reasoning is as follows. According to the Feynman-Stueckelberg interpretation, antiparticles are viewed as negative energy modes of the quantum field that propagate backward in time. An hypothetical  particle (not necessarily an anti-fermion or an anti-boson) that propagates backward in time would have transition rates that, instead of depending on the number of particles in the destination level, depend on the number in the origin level, $\bar{n}_1$. The term that depends on $n_2$ in \eqref{eq:w} must be constant. The possibility of time reversed particles is mentioned here as a motivation to consider transition rates in which the roles of origin and destination levels are interchanged. The conjecture that there are particles that meet this condition is proposed without assuming that they are particles or antiparticles. If that is the case, statistics should satisfy the condition
\begin{equation}\label{eq:const2}
\frac{e^{-\beta \mu_{\text{q}}/2}}{(1 + \beta \bar{n} \mu_{\text{q}}')^{1/2}} = 1,
\end{equation}
where subindex 2 in $\bar{n}_2$ was removed for simplicity. The only difference with Eq.\ \eqref{eq:const} is that the exponential has the opposite sign. Solutions are
\begin{equation}\label{eq:sol2}
e^{-\beta \mu_{\text{q}}} = \frac{1}{1 - \kappa/\bar{n}},
\end{equation}
with $\kappa$ a constant. Using this solution for $e^{-\beta \mu_{\text{q}}}$ in the expression for the mean number of particles, Eq.\ \eqref{eq:barn}, we obtain
\begin{equation}\label{eq:exotic}
\bar{n}/g = e^{-\beta(\epsilon-\mu)} + \kappa/g.
\end{equation}
As before, $\kappa=0$ corresponds to classical particles. The other case that is considered here is $\kappa=g$, that corresponds to the so-called statistics of ewkons \cite{hoyuelos-sisterna,hoyuelos1,hoyuelos2}. Calling $\bar{n}_\text{ewk}=\bar{n}/g$ the average number of ewkons in one of the $g$ states, then 
\begin{equation}\label{e.newk}
\bar{n}_\text{ewk} = e^{-\beta(\epsilon-\mu)} + 1.
\end{equation}

What are the features of particles that produce ewkon statistics? This question is addressed in the next subsection.

\subsection{Features of ewkons}
\label{s.features}

What happens when particles are identical but not fully distinguishable or indistinguishable? In this subsection, it is shown that an intermediate category of particles, that can be called sub-distinguishable, has ewkon statistics. In order to do that, the basic framework is presented first.

There are $N$ particles in $L$ states; each state has energy $\epsilon_i$ with $i=1\cdots L$. The Hamiltonian of non-interacting particles is $\hat{H} = \sum_{i=1}^L \hat{n}_i \epsilon_i$ where $\hat{n}_i$ is the particle number operator of state $i$. Degeneracy is taken into account implicitly since the values of $\epsilon_i$ may be the same for different $i$ (index $i$ represents a set of quantum numbers that determines the state). Note that this definition of number $n_i$ is different from the one in the previous sections; it does not encompass the number of particles in degenerate states or in close energy levels when the volume is large, it is an intensive quantity the mean value of which is the average number of particles of the previous sections divided by the corresponding $g_i$.  

Using the eigenstates of the number operators, the system's state is given by
\begin{equation}\label{e.nnn}
|n_1 \cdots n_L\rangle = |n_1\rangle\cdots |n_L\rangle.
\end{equation}
As mentioned in Sec.\ \ref{s.basic}, the canonical partition function is
\begin{equation}\label{e.ZN}
\mathcal{Z}_N = {\sum_{\{n_i\}}}' \langle n_1 \cdots n_L| e^{-\beta\hat{H}} |n_1 \cdots n_L\rangle = {\sum_{\{n_i\}}}' \prod_i \mathcal{Z}_{n_i}.
\end{equation}
Including a super-index $\mp$ to indicate symmetric or anti-symmetric states, the partition function of state $i$ with $n_i$ particles is
\begin{equation}\label{e.Zn}
\mathcal{Z}_{n}^\mp = {}^\mp\langle n|e^{-\beta\hat{n}\epsilon}|n\rangle^\mp = e^{-\beta n\epsilon} w^\mp
\end{equation}
where $w^\mp$ is the statistical weight factor; $w^+=1$ for bosons and $w^-=\delta_{n,0}+\delta_{n,1}$ for fermions; sub-index $i$ was omitted for simplicity. Let us call $|i_j\rangle$ the normalized state of particle $j$. The $n$ non-interacting particles are described by the number sate $|n\rangle$ that, in turn, is given by the symmetrization or anti-symmetrization of one-particle product states:
\begin{equation}\label{e.number}
|n\rangle^\mp = \frac{1}{\sqrt{\mathcal{N}}} \sum_{P} (\mp 1)^p P|i_1\cdots i_n \rangle,
\end{equation} 
where the sum is over possible permutations $P$, $\mathcal{N}$ is the normalization factor and $p$ is the parity of permutation $P$, with $P|i_1\cdots i_n \rangle=|i_{P1}\cdots i_{Pn} \rangle$. Since the $n$ particles are in the same state, all permutations are equal. The normalization is $\mathcal{N}=n!^2$ for bosons or fermions (in the last case the normalization is 1 since $n$ takes values 0 or 1). 

Instead, for classical distinguishable particles, the statistical wight factor is
\begin{equation}\label{e.wdist}
w_\text{dist} = 1/n!,
\end{equation}
an expression that leads to the Maxwell-Boltzmann distribution. There are several possible justifications that include the assumption that identical classical particles should be treated as permutable; more extensive discussions on this fundamental subject can be found in, for example, \cite{saunders,dieks,darrigol,saunders2}. In any case, the statistical weight factor for classical particles is 1 over the number of distinguishable configurations, $n!$.

There are two extreme situations: all particles indistinguishable with one possible configuration, or all particles distinguishable with $n!$ possible configurations. In the first case, $|i_{P1}\cdots i_{Pn} \rangle = |i_1\cdots i_n \rangle$ holds for any $P$ because $i_1 = \cdots = i_n$. In the second case, the equality $\{i_{P1}\cdots i_{Pn} \} = \{i_1\cdots i_n \}$ holds only if $P$ is the identity. (Ket notation, $|i_1\cdots i_n \rangle$, is used here only for indistinguishable particles; in other cases, curly brackets are used.)

\begin{figure}
	\centering
	\includegraphics[width=\columnwidth]{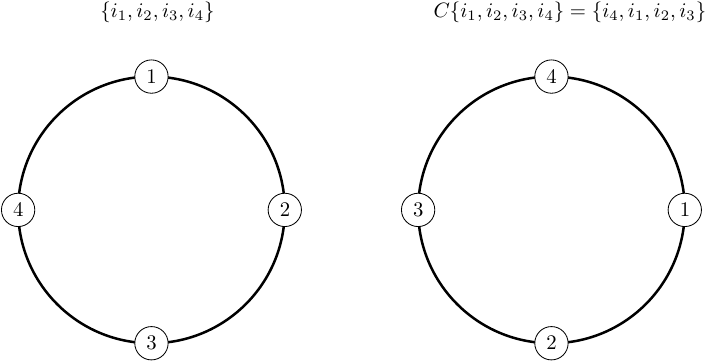}
	\caption{System of four particles arranged on a ring. A cyclic rotation generates an indistinguishable configuration (right) due to rotation symmetry. In cycle notation, operator $C$ is represented by $(1,4,3,2)$, meaning that element 1 goes to 4, 4 goes to 3, 3 goes to 2 and 2 goes to 1.}\label{f.cycle}
\end{figure}

The purpose is to analyze the possibility of an intermediate situation. Consider a system of $n$ particles that are arranged along a ring with rotation symmetry. The arrangement order can be distinguished, but cyclic rotations produce indistinguishable configurations, see Fig.\ \ref{f.cycle}. The number of distinguishable configurations is $(n-1)!$; since this number is reduced from $n!$ to $(n-1)!$, the term ``sub-distinguishable particles'' is proposed. The resulting statistical weight factor,
\begin{equation}\label{e.wewk}
w_\text{ewk} = 1/(n-1)!,
\end{equation}
corresponds to statistics of ewkons. The same factor is obtained not only for cyclic rotations but for any cyclic permutation $C$ that satisfies $C^n=I$, where $I$ is the identity \cite[p.\ 29]{gross}. More generally, sub-distinguishable particles satisfy, by definition, the following condition: if $n>0$ and if $P$ is equal to a cyclic permutation $C$ applied $j$ times, that is $P=C^j$ (with $j=1, \cdots, n$), then $\{i_{P1}\cdots i_{Pn} \} = \{i_1\cdots i_n \}$. In other words, configurations that differ by cyclic permutations are indistinguishable. The partition function,
\begin{equation}\label{eq:Zewk}
\mathcal{Z}_{n,\text{ewk}} = \frac{e^{-\beta \epsilon n}}{(n-1)!}
\end{equation}
generates the statistics of ewkons for identical sub-distinguishable particles. 

The (one-level) grand partition function for ewkons is 
\begin{align}\label{e.Q}
\mathcal{Q}_\text{ewk}&=\sum_n e^{\beta\mu n} \mathcal{Z}_n = \sum_{n=1}^\infty \frac{e^{-\beta (\epsilon-\mu)n}}{(n-1)!}= \sum_{m=0}^\infty \frac{e^{-\beta (\epsilon-\mu)(m+1)}}{m!} \nonumber \\
&= \exp[-\beta(\epsilon-\mu) + e^{-\beta (\epsilon -\mu)}]
\end{align}
and the average number of particles is given by Eq.\ \eqref{e.newk}, $\bar{n}_\text{ewk}=e^{-\beta (\epsilon -\mu)}+1$.  

From Eqs.\ \eqref{eq:barn} (with $\bar{n}_\text{ewk}=\bar{n}/g$) and \eqref{eq:muq}, the excess chemical potential is given by $e^{-\beta \mu_{\text{q}}} = \bar{n}_\text{ewk} e^{\beta (\epsilon-\mu)}$, so
\begin{equation}\label{e.muqewk}
e^{-\beta \mu_{\text{q}}} = \frac{1}{1 - 1/\bar{n}_\text{ewk}}.
\end{equation}
with the condition $\bar{n}_\text{ewk}>1$.

Using \eqref{e.muqewk} in Eq.\ \eqref{eq:w}, the transition rate from state 1 to state 2, with $\bar{n}_1$ and $\bar{n}_2$ average particle numbers (subscript `ewk' is removed) is
\begin{equation}\label{e.Wewk}
W_{\bar{n}_1,\bar{n}_2} = \nu e^{-\beta(\epsilon_2-\epsilon_1)/2} (1-1/\bar{n}_1).
\end{equation}
It depends only on the particle number in the origin level. If there is only one particle, it can not leave its state since the transition rate is zero (it is an absorbing state). Starting from an initial configuration in which $n_i\ge 1\; \forall i$, this condition remains valid for any subsequent time. If the initial configuration has empty states, it irreversibly evolves to the equilibrium configuration in which all states are occupied with at least one particle, with additional particles provided by the reservoir, that keeps the chemical potential $\mu$. The problem of a divergent total particle number (or total energy) is solved by including an upper limit for the energy, $\epsilon_\text{max}$ \cite[Sec.\ 5]{hoyuelos2}. The limit can be assumed as a feature of the density of states or can be a consequence of a limitation of the reservoir to provide additional particles (without the limit, the number of particles that the reservoir has to transfer to the system diverges).

The procedure includes the possibility of super-distinguishable particles, called genkons in \cite{hoyuelos-sisterna}; $n$ particles of this kind have $(n+1)!$ distinguishable configurations. This case corresponds to $\kappa=-g$ in \eqref{eq:exotic}, resulting in an average number of particles $\bar{n}_\text{gen}=e^{-\beta(\epsilon-\mu)}-1$. An interesting symmetry of ewkons and genkons is that they have average particle numbers equal to the inverse of those of fermions and bosons, respectively, if the sign of $\epsilon-\mu$ is inverted. Nevertheless, super-distinguishable particles are not further discussed because, since $n$ can be equal to $-1$, they lack, so far, a satisfactory physical interpretation.

\subsection{Ewkons and dark energy}
\label{s.darkenergy}

The equation of state parameter of a perfect fluid is a dimensionless number defined as
\begin{equation}\label{e.w}
w = \frac{P}{\rho}
\end{equation}
where $P$ is the pressure and $\rho$ is the energy density. It has been observed that the universe has a negative equation of state parameter $w$, close to $-1$. In Ref.\ \cite{Kowalski_2008} a value $w=-0.969$ has been reported, with statistical and systematic errors around $0.063$. The authors of Ref.\ \cite{Planck2015} established an upper bound $w<-0.94$ at $95\%$ confidence level. This negative value of $w$ cannot be produced by ordinary matter; it is related to the observed accelerated expansion of the universe and is mainly attributed to the presence of dark energy \cite{caldwell,Planck2015}.

An ideal ewkon gas is an appropriate candidate for the description of dark energy since it has an equation of state parameter $w$ close to $-1$ \cite{hoyuelos-sisterna}. Since the partition function for ewkons is available, pressure and energy density can be obtained using standard thermodynamic relations. As mentioned before, an upper limit for the energy, $\epsilon_\text{max}$, has to be taken into account.

It has been shown in Ref.\ \cite{hoyuelos1} that a nonrelativistic ideal gas of ewkons of mass $m$ and chemical potential $\mu$ has an equation of state parameter given by
\begin{equation}\label{e.wrel}
w_\text{ewk} = -1 + \frac{5}{3}\frac{\mu}{\epsilon_\text{max}},
\end{equation}
where it has been assumed that $\epsilon_\text{max}$ is much larger than $\mu$ or $k_B T$, and higher orders of $1/\epsilon_\text{max}$ are neglected.

Quantum field theory has been used in \cite{hoyuelos2} to analyze the case of a massless scalar field of ewkons, with $\mu=0$, and the following value for $w$ was obtained:
\begin{equation}\label{e.massless}
w_\text{ewk} = -1 + \frac{32}{\beta^4 \epsilon_\text{max}^4} \qquad (\mu=0),
\end{equation}
where, as before, $\epsilon_\text{max}\gg k_B T$, and higher order terms are neglected.

In both cases, a value of $w_\text{ewk}$ close to $-1$ is obtained for large enough values of $\epsilon_\text{max}$. This result suggests the possibility of using ewkons for the description of dark energy.

\subsection{Toy model}
\label{s.toy}

Simplified models are useful for a better understanding of different particle features. For example, particles with hard-core interaction in a one-dimensional lattice, with jump rates between neighboring sites that include an external force, reproduce the Fermi-Dirac statistics \cite{kania,suarez}. Hard-core interaction plays the role of the Pauli exclusion principle. An equivalent model for ewkons is proposed in the next paragraphs.

\begin{figure}
	\includegraphics[width=\columnwidth]{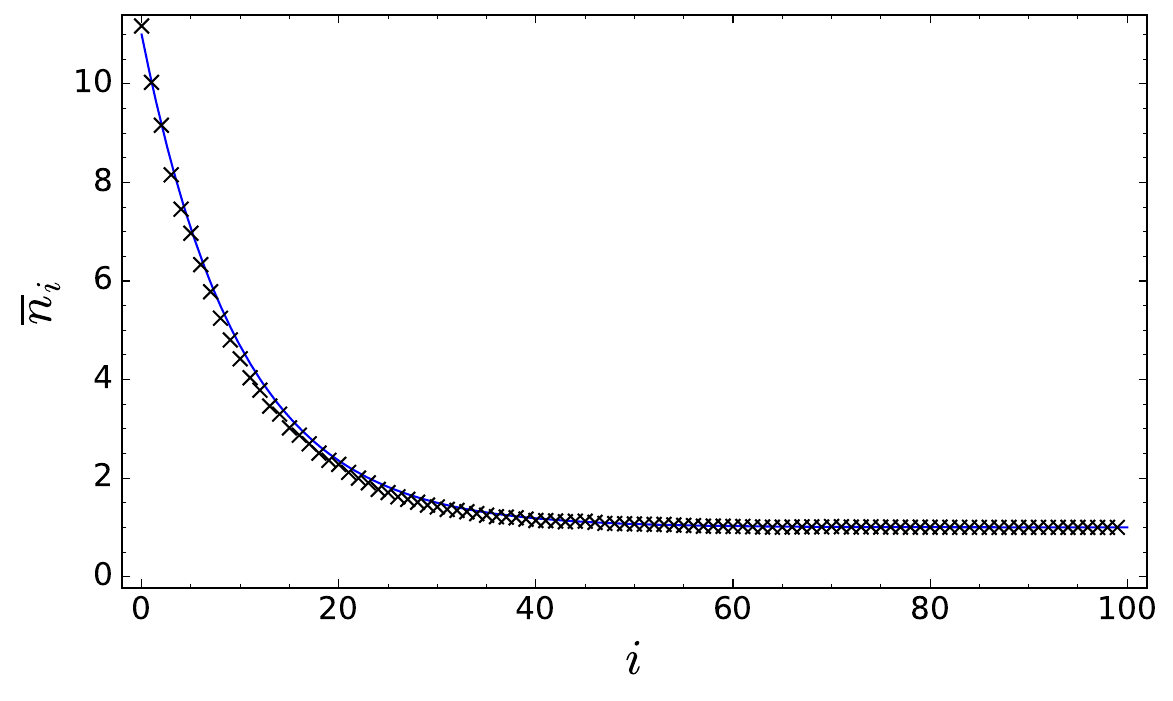}
	\caption{Average number of particles, $\bar{n}_i$, against lattice position $i$ with transition rates for ewkons, Eq.\ \eqref{e.trewk}. The initial condition is random. After a time of $15000$ MC steps the equilibrium distribution is reached. Crosses are averages taken during $5000$ MC steps. Parameters: $N=200$, $L=100$, $\beta \Delta\epsilon=0.1$. The blue curve is the equilibrium distribution for ewkons, Eq.\ \eqref{e.newk}.}\label{f.sim}
\end{figure}

A one-dimensional lattice of size $L$ represents a discrete energy space; the energy of site $i$ is $\epsilon_i = i\Delta \epsilon$, with constant energy gaps, $\Delta \epsilon = \epsilon_{i+1}-\epsilon_i\ge 0$, between any two neighboring sites. At $t=0$, $N$ particles are randomly distributed in the $L$ sites. A particle in site $i$ jumps to the right or to the left with the rates corresponding to ewkons:
\begin{align}\label{e.trewk}
W_\text{r} &= \nu e^{-\beta \Delta\epsilon/2} (1-1/n_i) \nonumber \\
W_\text{l} &= \nu e^{\beta \Delta\epsilon/2} (1-1/n_i)
\end{align}
Only jumps to nearest neighbors are allowed. In a small time interval $\Delta t = 1/(2 N \nu e^{\beta \Delta\epsilon/2})$, the jump probabilities are
\begin{align}
P_\text{r} &= e^{-\beta \Delta\epsilon} (1-1/n_i)/(2N) \nonumber \\
P_\text{l} &= (1-1/n_i)/(2N).
\end{align}
In a Monte Carlo (MC) simulation, a particle is randomly chosen with probability $1/N$ and then jumps to the right or to the left with probabilities $e^{-\beta \Delta\epsilon} (1-1/n_i)/2$ and $(1-1/n_i)/2$ respectively; a MC step is a time interval, equal to $N\Delta t$, in which each particle has, on average, one chance to jump. Like fermions, ewkons have an effective repulsive interaction. A particle can jump to another state only if there is another particle with it. The more particles are in the same state (or site), the more probable it is to jump. Figure \ref{f.sim} shows numerical results of $\bar{n}_i$ against position $i$, with zero current condition at the borders (see the figure caption for the simulation parameters). After a transient, the average number of particles converges to Eq.\ \eqref{e.newk}: $\bar{n}_\text{ewk} = e^{-\beta(\epsilon-\mu)} + 1$, represented by the blue curve. The value of $e^{-\beta\mu}$ is obtained from the condition $N = \sum_{i=1}^L \bar{n}_i$.

If $N<L$, the equilibrium distribution is not reached, and the final state is a frozen configuration, because there is zero or one particle in each site; the final configuration depends on the initial condition.

\section{Conclusions}
\label{s.conclusions}

The Widom insertion formula relates the excess chemical potential with the energy needed to insert one particle. The formula has been previously applied to derive a general expression of transition rates in a classical system of interacting particles \cite{dimuro}. These calculations are adapted for a quantum system of non-interacting particles at temperature $T$, for which the excess chemical potential, $\mu_{\text{q}}$, represents quantum effects (it is zero for the classical ideal mixture). From detailed balance, transition rates can be related to insertion energies that, in turn, are related to the excess chemical potential trough the Widom insertion formula. The result for the transition rate $W_{n_1,n_2}$, from energy level 1 to energy level 2, is given in terms of the energy difference, $\epsilon_2-\epsilon_1$, and the excess chemical potential in both levels, see Eq.\ \eqref{eq:w}. The expression evaluated at average particle numbers reproduces known results for fermions and bosons that include a factor $1\mp \bar{n}_2/g_2$ respectively. With this factor in the transition rates, the master equation (or Fokker-Planck equation if the energy space is continuous) gives an irreversible evolution towards the Fermi-Dirac or Bose-Einstein distributions. 

The expression obtained for transition rates provides an alternative procedure to derive known statistics. Starting from the hypothesis that the transition rate depends only on the number of particles in the destination level, then Fermi-Dirac, Bose-Einstein and Maxwell-Boltzmann statistics are obtained. This result prompts an exploration in a time-reversed situation,  where origin and destination levels are interchanged. The previous hypothesis now reads as follows: the transition rate depends only on the number of particles in the origin level (instead of destination level). Statistics of ewkons is derived from this hypothesis. 

The statistical weight factor for distinguishable particles is $1/n!$, while for ewkons it is $1/(n-1)!$. The number of distinguishable configurations for ewkons is reduced with respect to classical particles; for this reason, the term `sub-distingui\-shable particles' is also used. A sufficient condition to obtain ewkon statistics is to consider that particle configurations that differ by a cyclic permutation are indistinguishable; this condition reduces the number of distinguishable configurations to $(n-1)!$. Ewkon statistics, like Maxwell-Boltzmann statistics, is outside the scope of the spin-statistics theorem, since the theorem applies to indistinguishable particles. 

In summary, Widom's insertion formula provides useful information for determining transition rates in a quantum system of non-interacting particles. The form of the result suggest the possibility of sub-distinguishable particles with ewkon statistics when transition rates depend only on the number of particles in the origin level. It has been shown in previous works \cite{hoyuelos-sisterna,hoyuelos2} that an ideal gas of ewkons has a negative relation between pressure and energy density, a feature that makes them appropriate for the description of dark energy.

\section*{Acknowledgments}
This work was partially supported by Consejo Nacional de Investigaciones Cient\'ificas y T\'ecnicas (CONICET, Argentina, PIP 112 201501 00021 CO).

\section*{Appendix}

The demonstration that Fermi-Dirac and Bose-Einstein distributions are solutions of the master equation \eqref{e.master} in equilibrium is presented in this appendix. In equilibrium we have that $\frac{\partial \bar{n}_i}{\partial t}=0$ for all $i$. From Eq.\ \eqref{e.master}, a sufficient condition for equilibrium is that
\begin{equation}\label{e.equil}
\bar{n}_j W_{\bar{n}_j,\bar{n}_i} = \bar{n}_i W_{\bar{n}_i,\bar{n}_j}.
\end{equation}
Using the transition rates of Eq.\ \eqref{e.transrates}, and after some rearrangement, we have
\begin{equation}
\frac{\bar{n}_j\, e^{\beta \epsilon_j}}{1\mp \bar{n}_j} = \frac{\bar{n}_i\, e^{\beta \epsilon_i}}{1\mp \bar{n}_i} = \text{const.}
\end{equation}
Identifying the level independent constant in the right-hand side with $e^{\beta \mu}$, where $\mu$ is the chemical potential, we get Fermi-Dirac (plus sign) and Bose-Einstein (minus sign) distributions:
\begin{equation}\label{e.FDBE}
\bar{n}_i = \frac{1}{e^{\beta (\epsilon_i-\mu)} \pm 1}.
\end{equation}

\bibliographystyle{elsarticle-num}
\bibliography{rates.bib}

\end{document}